\documentclass[epjCONF]{svjour}
\usepackage{graphicx}
\usepackage[varg]{txfonts} 
\usepackage[latin1]{inputenc}
\session-title{2011 Hadron Collider Physics symposium}
\newcommand{\pt}{\ensuremath{p_{\mathrm{T}}}}
\newcommand{\kt}{\ensuremath{k_{\mathrm{T}}}}

\begin{document}
\title{High-\pt{} results from ALICE}
\author{M. van Leeuwen \thanks{\email{m.vanleeuwen1@uu.nl}} {\it for the ALICE collaboration}}
\institute{Nikhef, National Institute for Subatomic Physics and Institute for Subatomic Physics of Utrecht University, Utrecht, Netherlands}
\abstract{
We report recent results of high-\pt{} measurements in Pb--Pb
collisions at $\sqrt{s_{NN}}=2.76$ TeV by the ALICE experiment and
discuss the implications in terms of energy loss of energetic partons
in the strongly interaction medium formed in the collisions.
} 
\maketitle
\section{Introduction}
\label{intro}
High-energy collisions of heavy nuclei are used to study the
high-temperature states of strongly interacting matter and the
expected transition from confined matter to a deconfined Quark-Gluon
Plasma. Partons with high transverse momentum \pt{} are formed in hard
scatterings which happen early in the collision. The produced partons
then propagate through the hot and dense medium and lose
energy through interactions with the medium. Measurements of
high-\pt{} particle production are used to study the interactions
between fast partons and the medium and to determine the medium
properties using these interactions. Here we report a number of recent
high-\pt{} results from ALICE, the dedicated heavy-ion experiment at
the Large Hadron Collider (LHC), from Pb--Pb collisons with a
centre-of-mass energy $\sqrt{s_{NN}}=2.76$ TeV recorded during the
heavy ion run of the LHC in November 2010.

\section{Nuclear modification factor}
\begin{figure}
\includegraphics[width=0.49\textwidth]{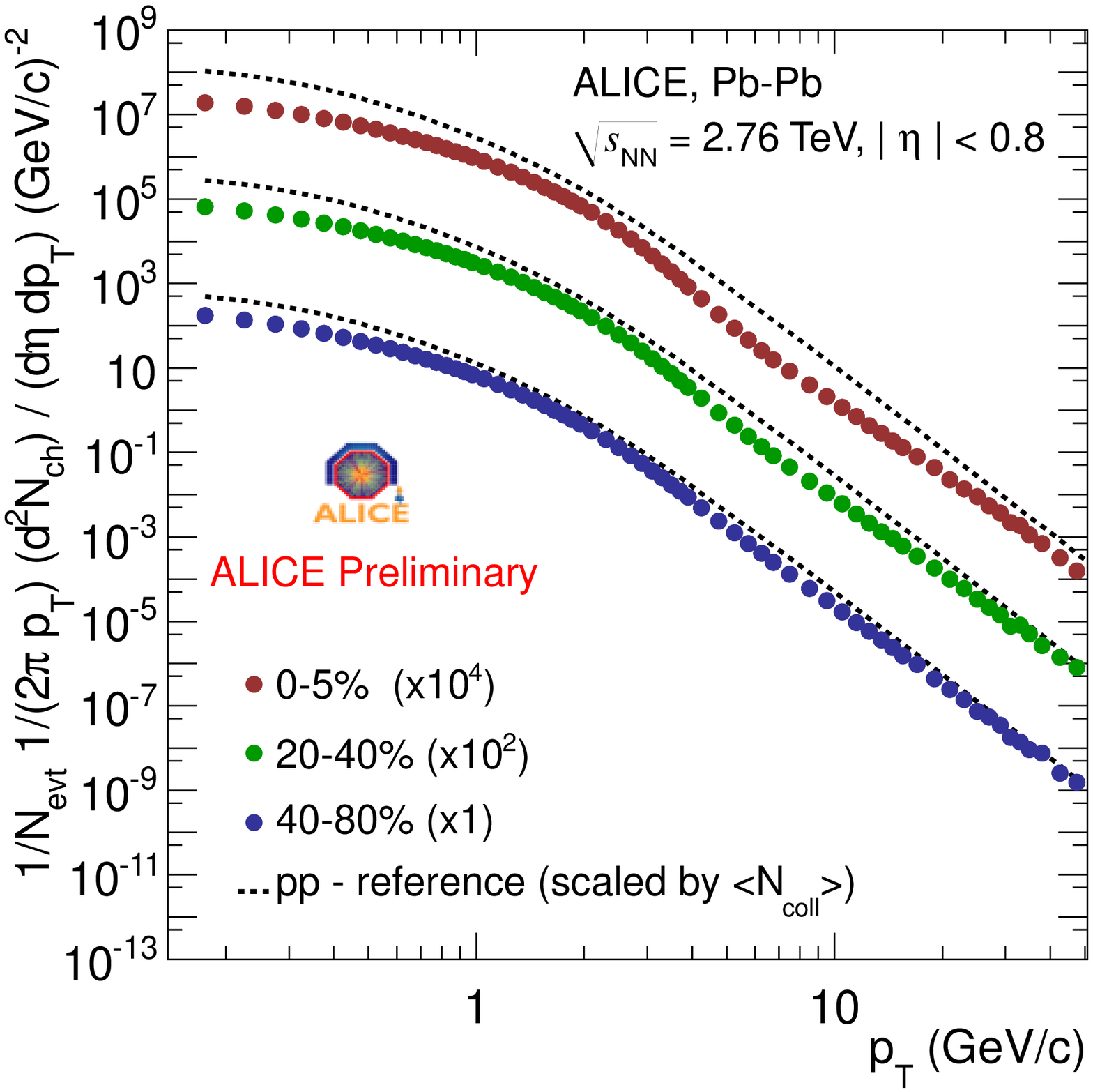}
\includegraphics[width=0.49\textwidth]{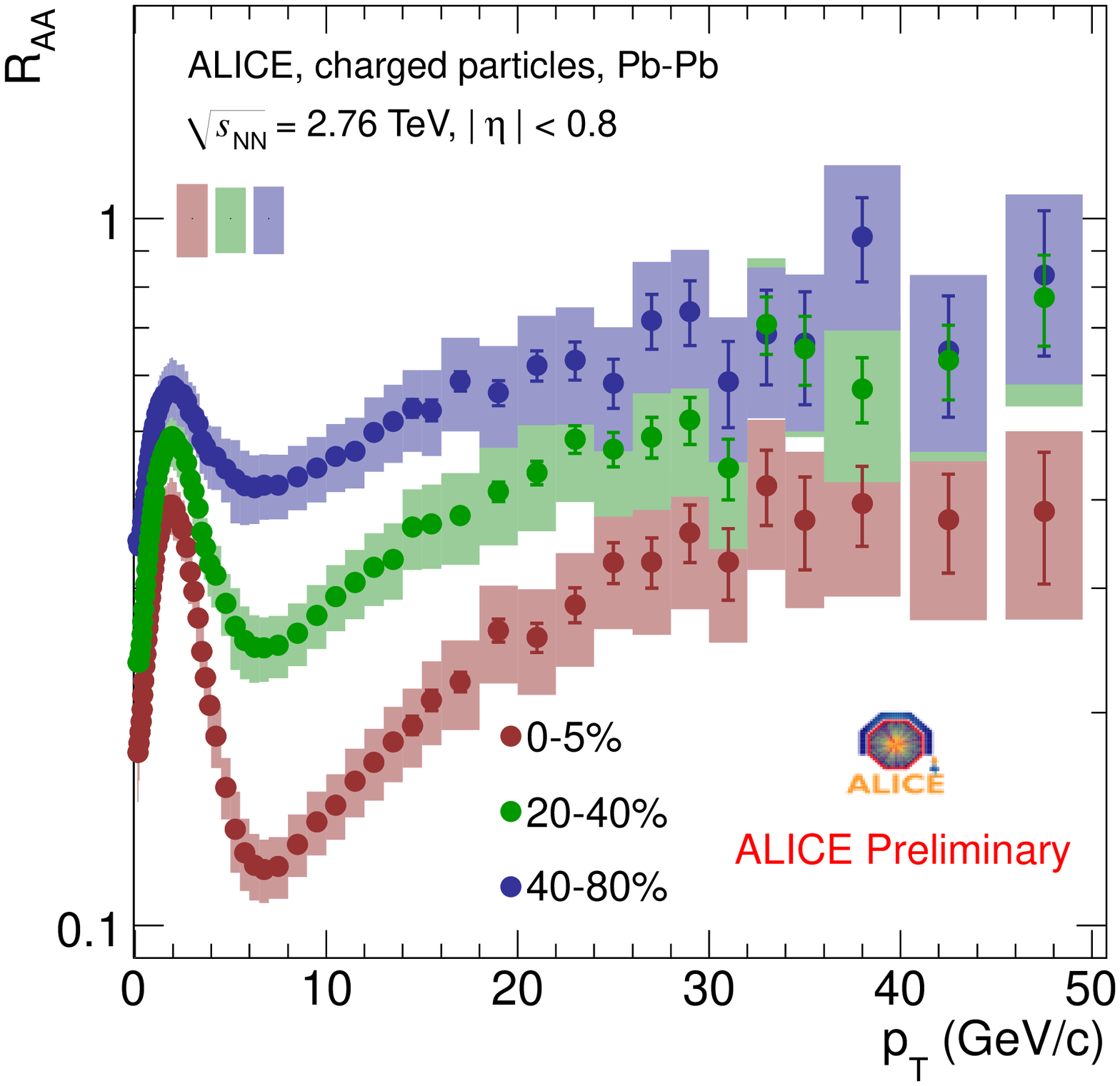}
\caption{\label{fig:spec_RAA}Left panel: Transverse momentum
  distributions of primary charged particles in Pb--Pb collisions at
  $\sqrt{s_{NN}}=2.76$ TeV with three different centrality
  selections. Right panel: nuclear modification factor $R_{AA}$ for
  charged particles in Pb--Pb collisions at
  $\sqrt{s_{NN}}=2.76$ TeV with three different centrality
  selections.}

\end{figure}

One of the most basic measurements that is sensitive to parton energy
loss in the hot and dense QCD medium are the charged particle production
spectra at high \pt{}. The measured transverse momentum spectra of
primary charged particles in
Pb--Pb collisions with three different
centrality selections are shown in the left panel of
Fig. \ref{fig:spec_RAA}. The collision centrality is determined using
the total multiplicity detected in the forward VZERO detectors and
reported as a fraction of the total hadronic cross section, with 0\%
labeling the most central events. The dashed lines in the Figure
indicate a parametrisation of the spectrum measured in pp
collisions, scaled by the total number of binary nucleon-nucleon
collisions $\langle N_\mathrm{coll} \rangle$ as determined from a
Glauber model \cite{Miller:2007ri,Aamodt:2010cz}. It can be seen in the figure that for
central collisions, the shape of the \pt-spectra in Pb--Pb collisions
is different from pp collisions, with a large suppression for
$\pt=4-10$ GeV/$c$.

To quantify the differences, the nuclear modification factor
\[
R_{AA} = \frac{dN/d\pt|_\mathrm{PbPb}}{\langle N_\mathrm{coll} \rangle dN/d\pt|_\mathrm{pp}},
\]
i.e. the ratio between the \pt-distributions
in Pb--Pb collisions $dN/d\pt|_\mathrm{PbPb}$ and in pp collisions
$dN/d\pt|_\mathrm{pp}$, scaled with the number of binary collisions, is
calculated. The nuclear modification factor for charged particles in
Pb--Pb collisions with three different centrality selections
is shown in the right panel of Fig. \ref{fig:spec_RAA}. The figure
clearly shows a significant suppression $R_{AA} < 1$. The effect is
largest for the most central bin $0-5\%$, where the medium density and
the average path length through the medium are the largest. The
strongest suppression is seen for $\pt \approx 7$ GeV/$c$, with a
gradual rise of $R_{AA}$ towards larger \pt. The increase of $R_{AA}$
with \pt{} is qualitatively consistent with the expectation that
parton energy loss $\Delta E$ is only weakly dependent on the parton
energy $E$, leading to a
decrease of the relative energy loss $\Delta E/E$ with increasing \pt.

\begin{figure}
\centering
\includegraphics[width=0.5\textwidth]{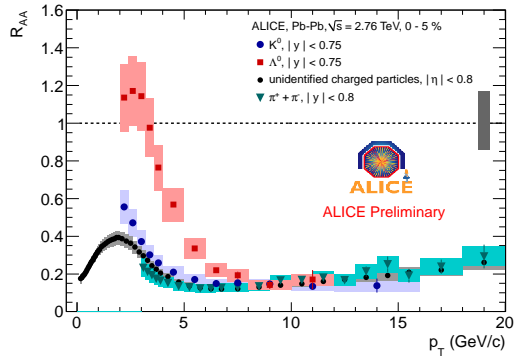}
\caption{Nuclear modification factor $R_{AA}$ for identified hadrons in
the 0-5\% most central Pb--Pb collisions at $\sqrt{s_{NN}}$=2.76 TeV.}
\label{fig:RAA_lamka}
\end{figure}
Figure \ref{fig:RAA_lamka} shows a comparison of the nuclear
modification factor for $\Lambda$ and $K_0^s$, measured using
reconstruction of the weak-decay topology, to the result for
unidentified charged particles and identified pions for the most
central events.  It is interesting to see that the $R_{AA}$ for
identified mesons shows a similar \pt-dependence to the charged
particles, while the $\Lambda$ show much smaller suppression at
intermediate $\pt < 6$ GeV/$c$. The enhancement of baryon production
compared to meson production at intermediate \pt, might be due to a
large contribution of hadron formation by coalescence of quarks from
the hot and dense medium \cite{Lin:2002rw,Hwa:2002tu,Fries:2003vb}.

At higher \pt, all hadrons show 
the same suppression, which suggests that the dominant energy loss
mechanism is at work at the partonic level. If hadronic energy loss would
be important, one would expect to see that different hadrons would
have different cross sections for the relevant energy loss mechanism
and thus be affected differently.


\section{Di-hadron measurements}
\begin{figure}
\centering
\includegraphics[width=0.8\textwidth]{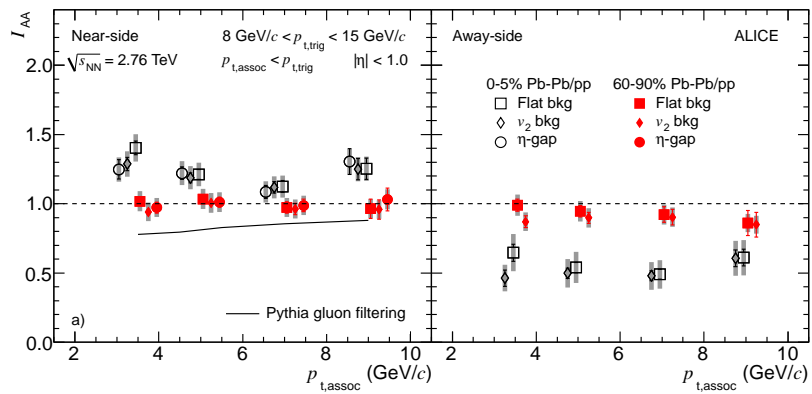}
\caption{\label{fig:alice_iaa}Ratios of measured charged hadron yield
  associated with a high-\pt{} {\it trigger} particle with $8 < \pt <
  15$ GeV/$c$ in Pb--Pb and pp collisions on the near
  ($|\Delta\phi|<0.7$, left panel)
  and away ($|\Delta\phi-\pi|<0.7$, right panel) sides, as a function
  of asociated particle \pt. Results are shown for two different centrality
  selections: peripheral 60-90\% (solid red data points) and central 0-5\%
  (open data points). Three different background subtraction methods
  are shown. The grey bars indicate systematic uncertainties from
  tracking efficiency and secondary particles \cite{ALICE:2011vg}.}
\end{figure}
Using di-hadron correlation techniques, the yield of charged particles
produced in association with high-\pt{} hadrons can be determined. In
these measurements, we distinguish the {\it near-side} yield of
particles produced in the same jet as the high-\pt{} {\it trigger}
hadron, and the {\it away-side} or recoil yield of particles in the
recoiling jet. This measurement has been performed in pp and Pb--Pb
collisions at $\sqrt{s_{NN}}=2.76$ TeV using a trigger particle
selection of $8 < \pt^\mathrm{trig} < 15$ GeV/$c$ and the ratio
$I_{AA}$ of the associated yield per trigger particle in Pb--Pb and pp collisions
is shown in Fig. \ref{fig:alice_iaa}. The results for peripheral
collisions (red points in Fig. \ref{fig:alice_iaa}) are very similar
to pp ($I_{AA} \approx 1$), while for central collisions (grey points)
a slight enhancement of the yield is seen on the near side and a
suppression on the away side. Both effects are qualitatively
consistent with expectations from parton energy loss in combination
with a trigger bias which cause the parton on the away-side to
typically have a larger energy loss than the one on the near side. The
enhancement on the near side suggests that the trigger particle selects
hard scattered partons with higher energy in the Pb--Pb collisions than in
pp collisions, due to energy loss of the leading parton.

\section{Constraining theoretical models}
\begin{figure}
\includegraphics[width=0.44\textwidth]{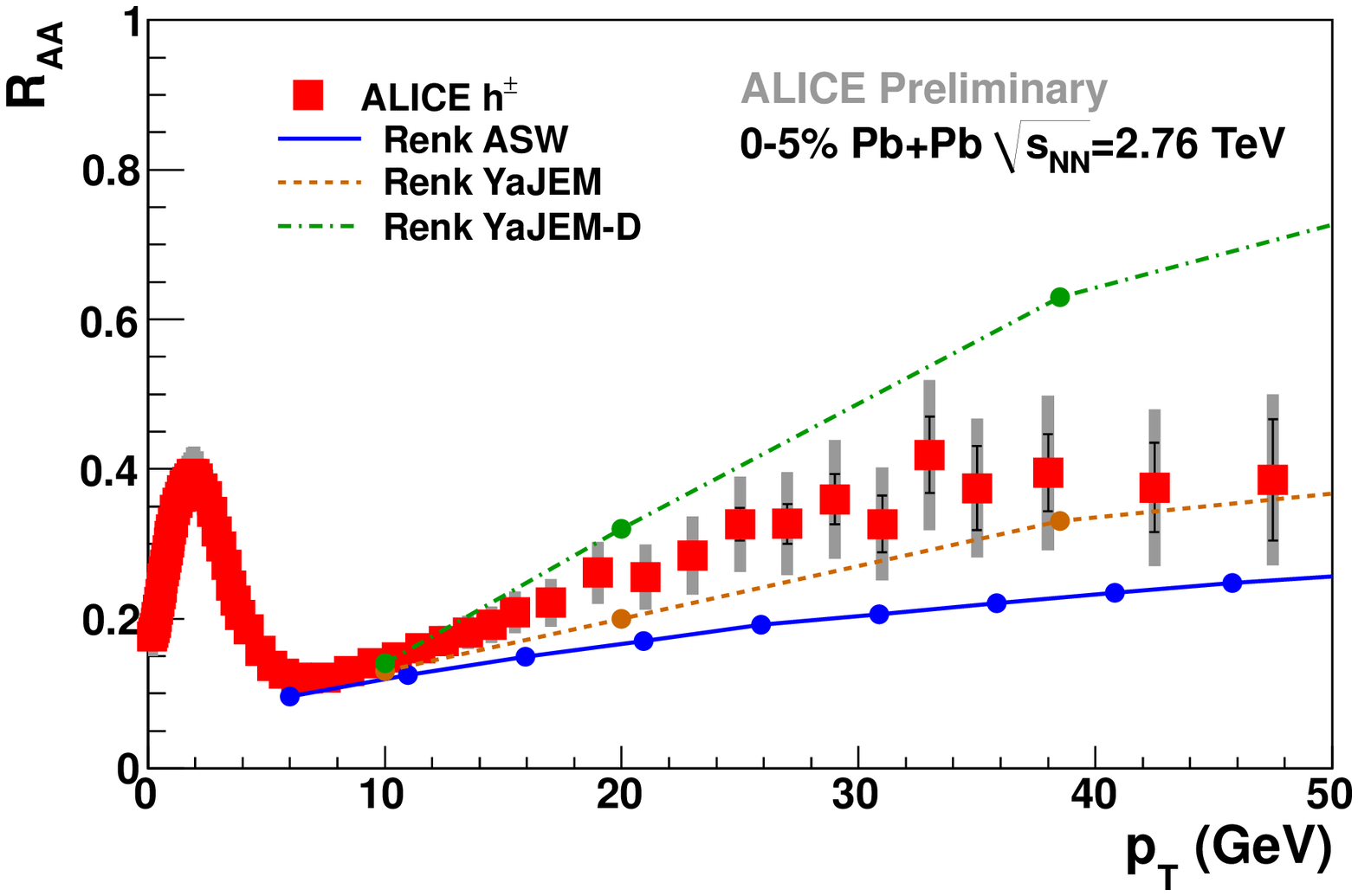}
\includegraphics[width=0.55\textwidth]{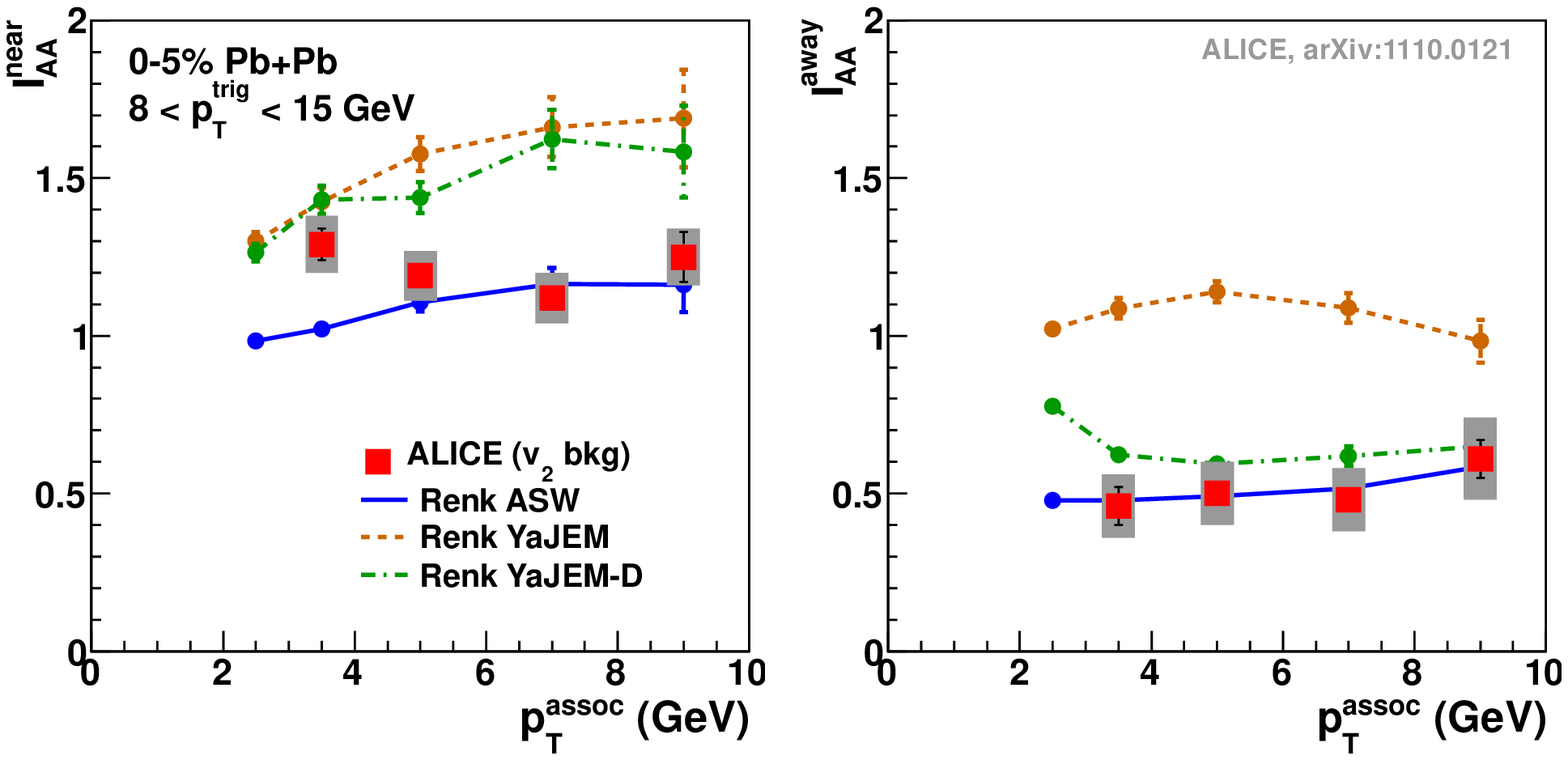}
\caption{\label{fig:RAA_IAA_renk}Nuclear modification factor $R_{AA}$
  and per-trigger associated yield modification factor $I_{AA}$ for
  the 0-5\% most central Pb--Pb collisions compared to model
  calculations (see text).}
\end{figure}
The single-inclusive nuclear modification factor $R_{AA}$ and the di-hadron
modification $I_{AA}$ sample the geometry of the collision zone with
different weights. A simultaneous comparison of both measurements with
theoretical calculations can be used to infer the path-length
dependence of energy loss
\cite{Renk:2007id}. Fig. \ref{fig:RAA_IAA_renk} shows a comparison of
the measured $R_{AA}$ and $I_{AA}$ to model calculations by Renk
\cite{Renk:2011wp,Renk:2011gj}.

The blue line in  Figure \ref{fig:RAA_IAA_renk} labeled `Renk ASW' indicates the expected
energy loss using the `quenching weights' calculation for the multiple
soft scattering approximation by Armesto, Salgado and Wiedemann
\cite{Salgado:2003gb} in a realistic medium-density profile, based on
hydrodynamical simulations. One overall scaling parameter was used to
relate the local transport coefficient $\hat{q}$ to the medium density
in the hydrodynamical model. This scaling parameter was tuned using
measurements from the Relativistic Heavy Ion Collider (RHIC) at
$\sqrt{s_{NN}}=200$ GeV. The fact that the blue line in the left panel
of Fig. \ref{fig:RAA_IAA_renk} is below the measured data points
indicates that the single hadron suppression increases less from RHIC
to LHC than expected based on the density inferred from multiplicity
measurements which are used to tune the hydrodynamical evolution. The
agreement of the ASW calculation with the di-hadron correlations is
better.

The other two lines in Figure  \ref{fig:RAA_IAA_renk} represent energy
loss calculations based on a Monte Carlo shower model YaJEM (`Yet
another Jet Energy-loss Model') \cite{Renk:2011wp,Renk:2011gj} in
which medium-induced radiation is generated by increasing the
virtuality as the parton propagates through the medium. Default YaJEM
(orange dotted lines in Fig. \ref{fig:RAA_IAA_renk}) agrees rather
well with the $R_{AA}$ measurement, while the di-hadron suppression is
too weak, due to the approximately linear dependence of the energy
loss on the path length $L$. YaJEM-D (green dashed lines in
Fig. \ref{fig:RAA_IAA_renk}) is identical to YaJEM, but
introduces a minimum virtuality of the parton $Q_0=\sqrt{E/L}$, related to the
requirement that the formation time of each in-medium shower is
shorter than its path length through the medium. This causes a
stronger path length dependence of the energy loss, similar to the
expected $L^2$ dependence due to the Landau-Pomeranchuk-Migdal
effect, which leads to a stronger suppression of the di-hadron recoil
yield. However, it should be noted that for the current setting of the
model, the inclusive hadron suppression is also smaller ($R_{AA}$
larger) than measured. Increasing the medium density in YaJEM-D would 
improve the agreement with the measurements.

All of the models presented in Figure \ref{fig:RAA_IAA_renk} show
deviations from the measured values in several places. A more
systematic comparison of models with the measurements will be needed
to quantify deviations of the models from the data and to disentangle
effects from the medium geometry and the path-length dependence of the
energy loss process itself.

\section{Heavy flavour}
\begin{figure}
\centering
\includegraphics[width=0.8\textwidth]{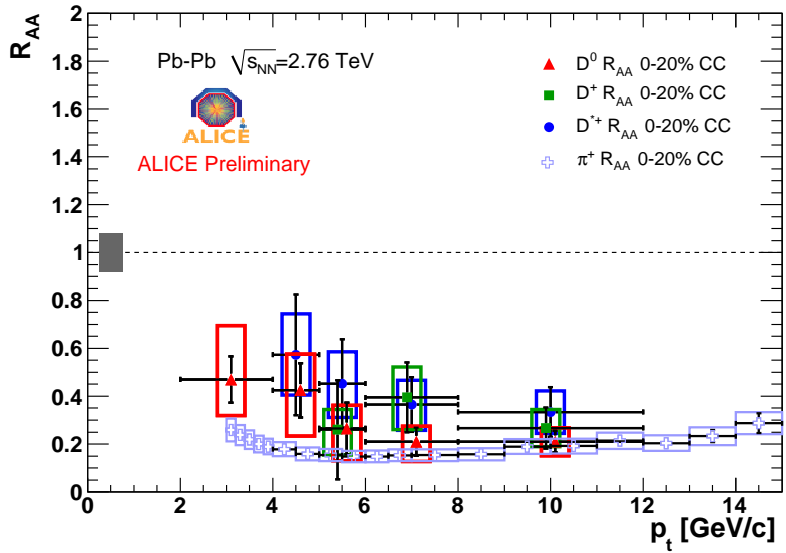}
\caption{\label{fig:RAAcharm}$R_{AA}$ for $D$ mesons in 0-20\% central
  Pb--Pb collisions at $\sqrt{s_{NN}}=2.76$ TeV. The $D$ mesons
are reconstructed from their hadronic decays: $D^0 \rightarrow K\pi$,
$D^{\pm} \rightarrow K\pi\pi$ and $D^{*\pm} \rightarrow \pi^{\pm} +
D^0 \rightarrow \pi^{\pm} + K\pi$.}
\end{figure}
A specific expectation for radiative parton energy loss is that the
effect will be smaller for heavy quarks at lower \pt, when the quarks
travel in the medium at speeds significanty below the speed of light,
due to the so-called `dead cone' effect
\cite{Dokshitzer:2001zm,Armesto:2003jh}. To test this expectation,
ALICE has measured the nuclear modification factor of $D$ mesons, as
shown in Fig. \ref{fig:RAAcharm}. The measured $D$ meson suppression
is slightly smaller than the values seen for pions, but the difference
is within the current statistical and systematic
uncertainties. Related measurements of muon and electron production
from heavy flavour decay are reported in \cite{suire}.

A more precise measurement using a larger data sample
and a careful comparison to theoretical expectations are needed to
determine whether the dead-cone effect is really observed in
experimental data. In addition, measurements of B mesons are planned,
which will have a larger discriminating power, because the dead-cone
effect is larger for the heavier $b$ quarks.

\section{Jets}
\begin{figure}
\includegraphics[width=0.54\textwidth]{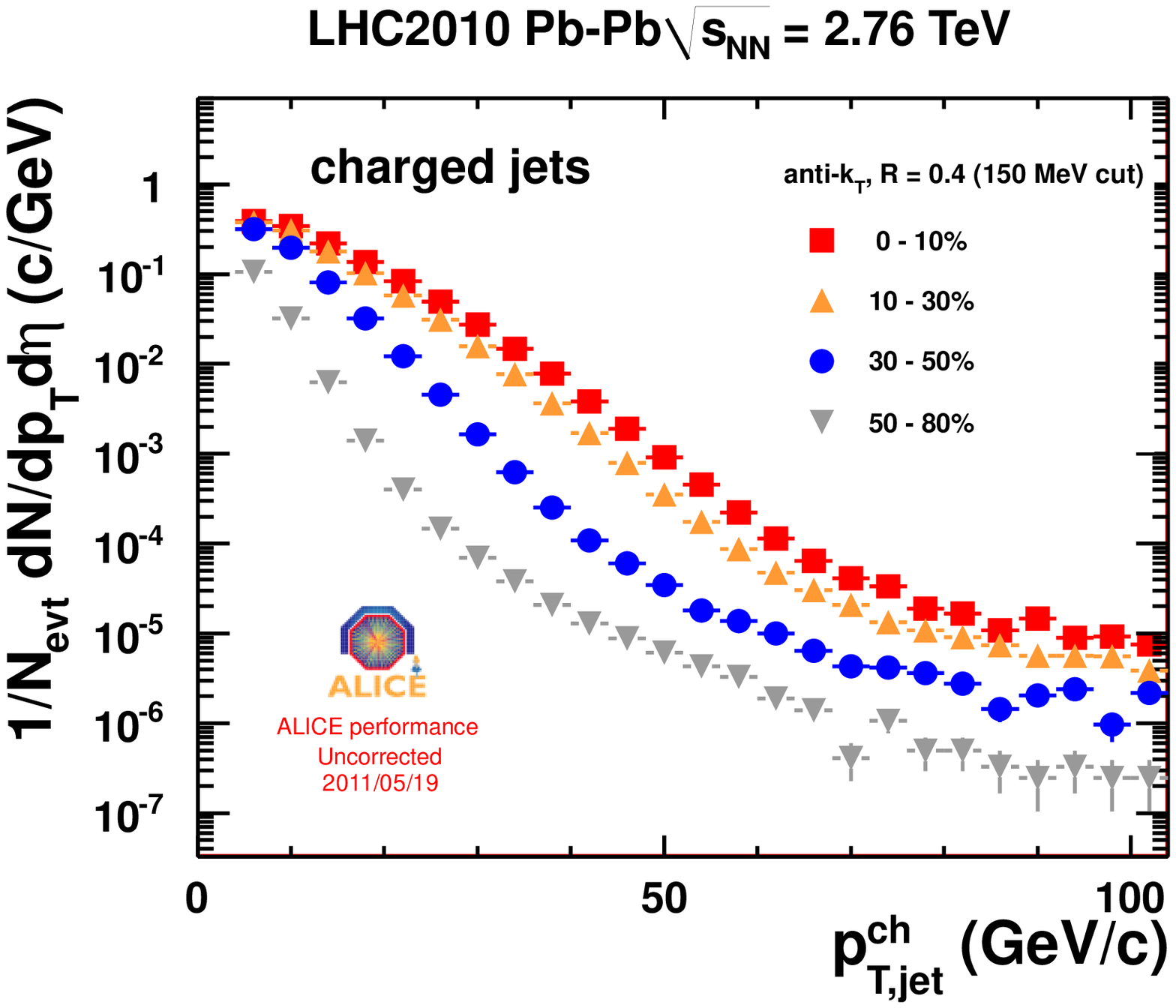}
\includegraphics[width=0.45\textwidth]{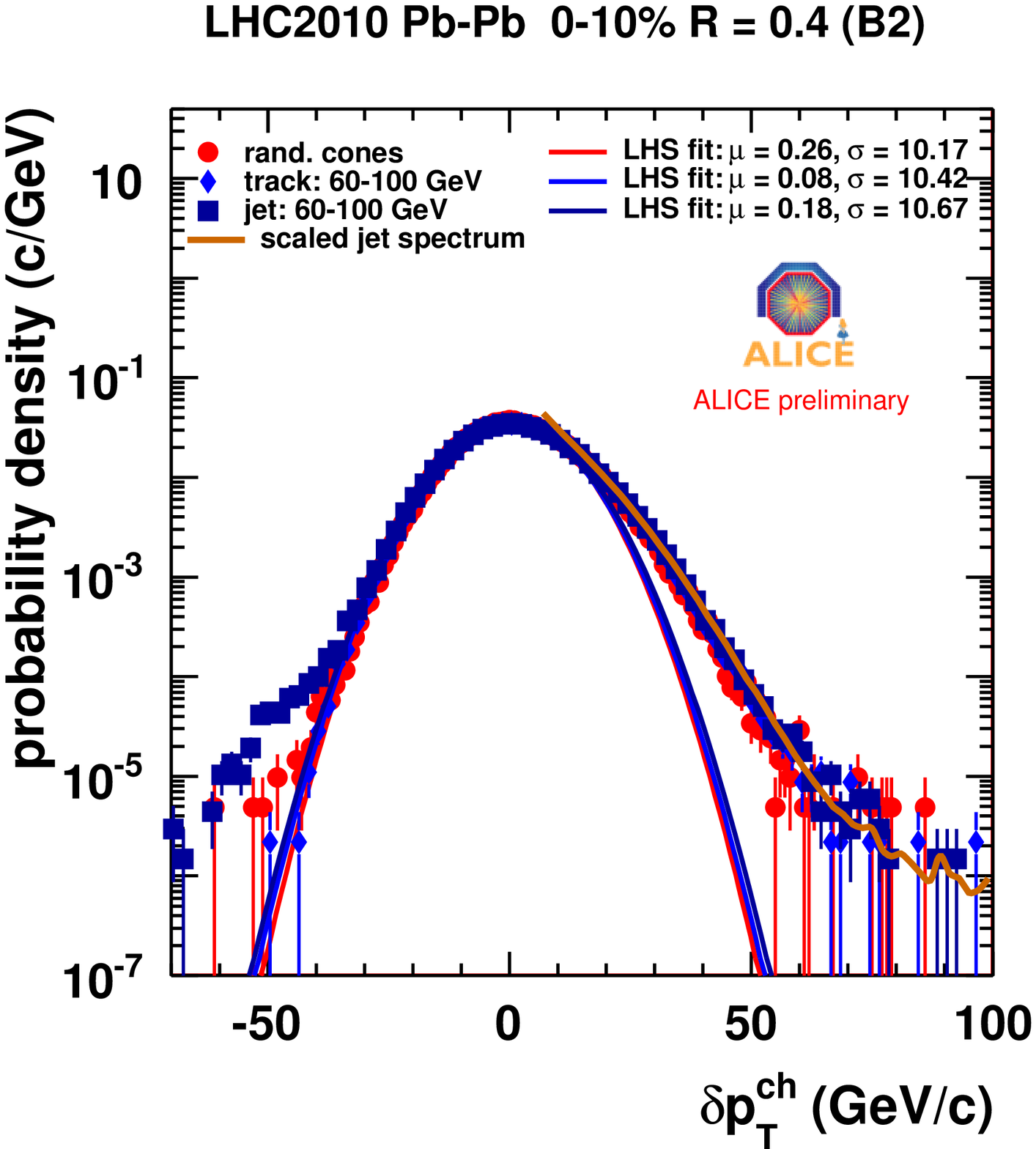}
\caption{\label{fig:jets}Left panel: Transverse momentum distributions
  of reconstructed jet spectra in Pb--Pb collisions with different
  centralities, using only charged particle tracks. The energy of the
  uncorrelated background has been subtracted, but no corrections are
  applied for background fluctuations and detector effects.  Right
  panel: Background energy distribution in central 0-10\% events,
  determined using random cones and jet and track embedding. The
  curves show Gaussian fits to the left-hand side (LHS) of the
  distributions.}
\end{figure}
Measurements of inclusive hadrons and di-hadrons at high \pt{} are
mostly sensitive to leading jet fragments, because the steeply falling
\pt-spectrum causes subleading
fragments to be overwhelmed by the much larger yields of fragments of
lower momentum partons. In addition, the measurements integrate over a
large range of energies of the orginal partons. Jet reconstruction in
Pb--Pb collisons has the potential to largely overcome both drawbacks:
if the jet cone radius is large compared to the typical angles of
gluon emission, most the radiated energy is recovered by the
jet-finding algorithm, which then provides a measure of the energy of
the parton from the hard scattering (before energy loss).

However, jet reconstruction in Pb--Pb collisions at the LHC is
challenging, due to the large underlying event energy. The left panel
of Fig. \ref{fig:jets} shows the reconstructed momentum distribution
of charged jets from the anti-\kt{} algorithm with $R=0.4$, after
subtraction of the uncorrelated background, which is measured on
event-by-event basis using the \kt{} algorithm from the FastJet
package \cite{Cacciari:2011ma}. At low \pt, a clear excess is visible
in the jet spectrum of central events compared to peripheral events
due to background fluctuations, which lead to `fake jets'. Judging
from the curvature of the jet spectrum, fluctuations/fake jets
dominate the jet spectrum up to $\pt \approx 70$ GeV/$c$ for central
(0-10\%) events.

The right panel of Fig. \ref{fig:jets} shows the background
fluctuations as measured directly using random cones and two types of
embedding \cite{ALICE:2012ej}. The random cone technique places `jet' cones
in the event at random location and then calculates the
background-subtracted transverse momentum in the cone to measure the
fluctuations. The embedding technique adds a track or several tracks
from a jet to the event and then compares the reconstructed transverse
momentum to
the transverse momentum of the input track or jet to measure the fluctuations. The
three methods give very similar results, indicating that the
background measurement is mostly sensitive to the track density (and
correlations) in the
event and not to details of the jet fragmentation or the placement of
the cone.

The jet results can only be interpreted in conjunction with the
background fluctuation measurement. ALICE is currently pursuing
unfolding techniques to remove the effect of the background
fluctuations from the reconstructed jet spectrum. Given the large
\pt-reach of the fluctuations, a larger data set is likely needed to
extend the measured jet spectrum and allow unfolding of the background
fluctuations.

\section{Outlook}
In these proceedings, we have reported first results on high-\pt{}
measurements of Pb--Pb collisions at $\sqrt{s_{NN}}=2.76$ TeV
performed by ALICE. The effects of the energy loss of partons
propagating through the hot and dense medium are clearly seen in the
suppression the inclusive yields of charged particles and
di-hadrons, as well as for heavy mesons.

The goal of this research is to develop a quantitative understanding
of the interactions between energetic partons and the medium and to
use this understanding to determine properties of the medium such as
the energy density or transport coefficient(s). A careful comparison
of multiple measurements with theoretical expectations is needed to
develop our understanding of parton energy loss. A first attempt of
such a comparison for the LHC energy was shown in
Fig. \ref{fig:RAA_IAA_renk}, but a more systematic approach will be
pursued in the near future.

The results in these proceedings are based on the data sample of about
20M hadronic interactions collected in the heavy ion run of the LHC in 2010. A
much larger data sample has been collected in November 2011, which
will allow to improve the precision of the heavy flavour and jet
measurements and extend the \pt-range over which the various
measurements can be performed. These improvements will be essential to
further constrain the theory of parton energy loss.

\bibliography{alice_highpt_mvanleeuwen_hcp}


\end{document}